\pdfoutput=1

\documentclass[11pt,twoside,a4paper,cmspaper,final,collab]{cms-tdr}
\def\svnVersion{238804}\def\cmsCernNo{2013-226}\def\cmsCernDate{\today}\def\cmsMessage{Published in Physical Review Letters as \href{http://dx.doi.org/10.1103/PhysRevLett.112.161802}{\doi{10.1103/PhysRevLett.112.161802.}}}

\begin{document}\cmsNoteHeader{SUS-13-014}

\hyphenation{had-ron-i-za-tion}
\hyphenation{cal-or-i-me-ter}
\hyphenation{de-vices}

\RCS$Revision: 223886 $
\RCS$HeadURL: svn+ssh://svn.cern.ch/reps/tdr2/papers/SUS-13-014/trunk/SUS-13-014.tex $
\RCS$Id: SUS-13-014.tex 223886 2014-01-21 05:46:58Z abarker $
\newlength\cmsFigWidth
\ifthenelse{\boolean{cms@external}}{\setlength\cmsFigWidth{0.85\columnwidth}}{\setlength\cmsFigWidth{0.5\textwidth}}
\ifthenelse{\boolean{cms@external}}{\providecommand{\cmsLeft}{top}}{\providecommand{\cmsLeft}{left}}
\ifthenelse{\boolean{cms@external}}{\providecommand{\cmsRight}{bottom}}{\providecommand{\cmsRight}{right}}
\ifthenelse{\boolean{cms@external}}
{\providecommand{\suppMaterial}{the supplemental material [URL will be inserted by publisher]}}
 {\providecommand{\suppMaterial}{the Appendix}}

\providecommand{\conep}{\widetilde{\chi}^{+}_{1}}

\cmsNoteHeader{SUS-13-014} 
\title{Search for top squark and higgsino production using diphoton Higgs boson decays}

\date{\today}

\abstract{
Results are presented of a search for a ``natural'' supersymmetry scenario with gauge mediated symmetry breaking.
It is assumed that only the supersymmetric partners of the top quark (the top squark)
and the Higgs boson (higgsino) are accessible.
Events are examined in which there are two photons forming a Higgs boson candidate, and at least two $\cPqb$-quark
jets. In 19.7\fbinv of proton-proton collision data at $\sqrt{s} = 8\TeV$, recorded in the CMS experiment,
no evidence of a signal is found and lower limits at the 95\% confidence level are set,
excluding the top squark mass below 360 to 410\GeV, depending on the higgsino mass.
}

\hypersetup{%
pdfauthor={CMS Collaboration},%
pdftitle={Search for top squark and higgsino production using diphoton Higgs boson decays},
pdfsubject={CMS},%
pdfkeywords={CMS, physics, SUSY}}

\maketitle 

The recent measurements of the properties of the new boson with mass around 126\GeV by ATLAS~\cite{atlas-hcoup, atlas-hspin}
and CMS~\cite{PhysRevLett.110.081803, cms-h} experiments confirm that it is consistent with the standard model (SM) Higgs boson.
One of the most important problems in particle physics is now to understand why
 the boson's mass is so small compared to the Planck scale.
Supersymmetry (SUSY) offers an elegant solution to this hierarchy problem.
Although experimental searches have found no signs of it so far
and the simplest SUSY models have become increasingly more constrained,
there remain very large yet unprobed areas of SUSY parameter space that can still be ``natural",
\ie have a fairly small amount of fine tuning.

The essential requirement of ``natural" SUSY is that the masses of the superpartners
of the top quark and the Higgs bosons, the top squark and the higgsinos,
be light (see, \eg \cite{susy-natural-1, susy-natural-2}).
This Letter describes a search we performed for events with a topology motivated by
 SUSY with a gauge mediated SUSY breaking (GMSB) model~\cite{gmsb-1, gmsb-kribs, gmsb-2, gmsb-3}.
Complete GMSB models are considered with a minimal number of accessible SUSY partners
to satisfy the naturalness requirements, namely the right-handed top squark and higgsinos.

The lightest chargino ($\conep$) and neutralinos
($\PSGczDo,\PSGczDt$) in this case are almost pure higgsinos and are therefore almost mass-degenerate.

Pairs of higgsinos are produced either directly through electroweak production or
through right-handed top squark-antisquark pairs (strong production) with cascade decays~
\begin{equation}
\stR \to \cPqb \conep \text{ or } \stR \to \cPqt \PSgxz ,
\end{equation}
where $i = 1, 2$.  The $\conep$ ($\PSGczDt$) subsequently decays into a very off-shell $W^+$ ($\Z$) boson and a $\PSGczDo$.
(Here, and in what follows, the charge-conjugate final states are also implied.)
The near mass degeneracy of the higgsino states requires the off-shell boson decay products to be very soft.

 The decay of the lightest neutralino in all cascades is~
\begin{equation}
\PSGczDo \to \PH \PXXSG \text{ or } \PSGczDo \to \Z \PXXSG ,
\end{equation}
where $\PXXSG$ is the gravitino. The branching fractions of $\PSGczDo$ decays depend on SUSY parameters.
For a significant portion of the parameter space, neutralino decays into Higgs bosons dominate.
This includes regions of low \tanb,
the ratio of the vacuum expectation values of the two SUSY Higgs doublets,
and negative values of the higgsino mass term
$\mu$~\cite{gmsb-shih}.

\begin{figure}[hbtp]
  \begin{center}
    \includegraphics[width=0.8\cmsFigWidth]{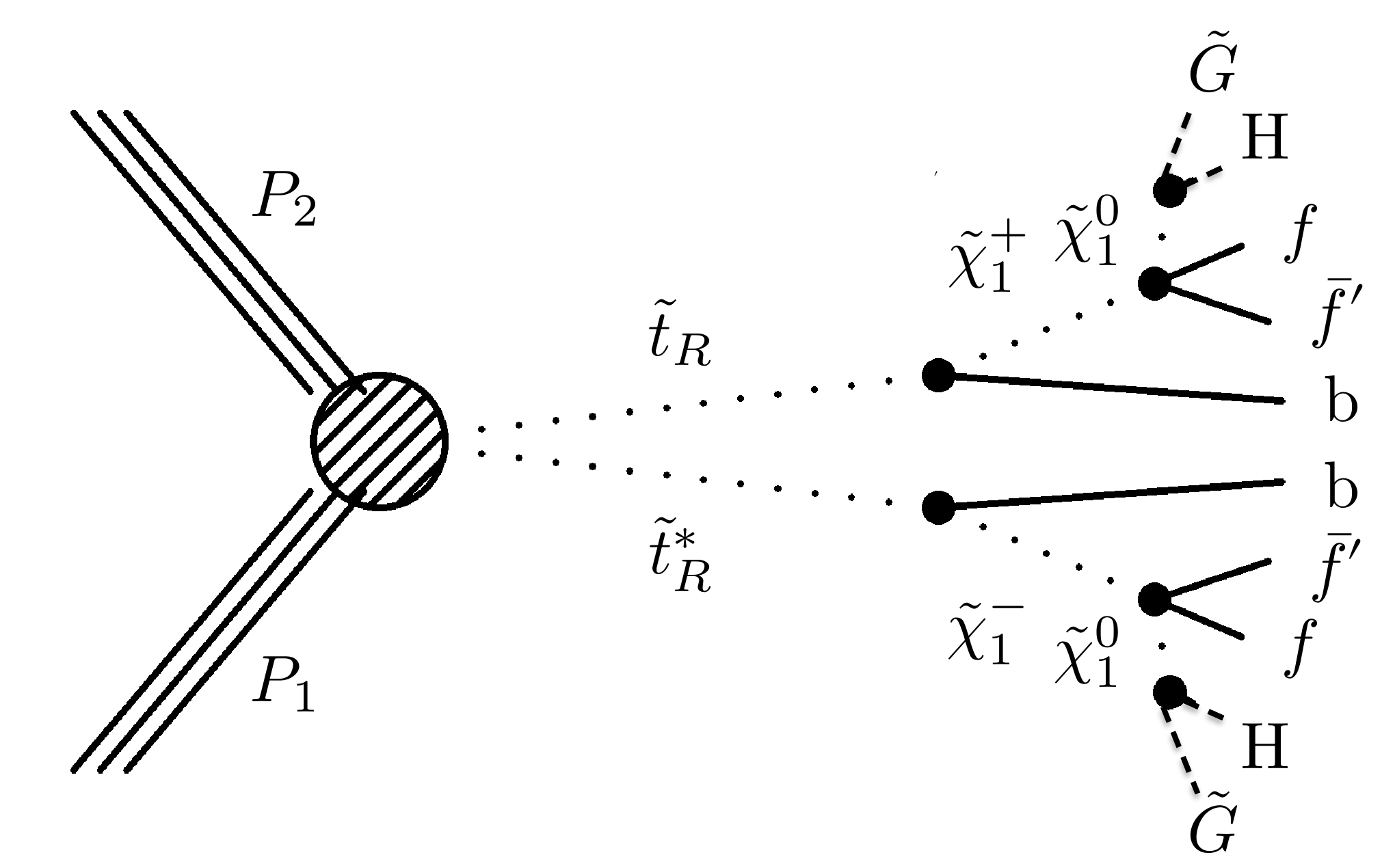}
    \includegraphics[width=0.8\cmsFigWidth]{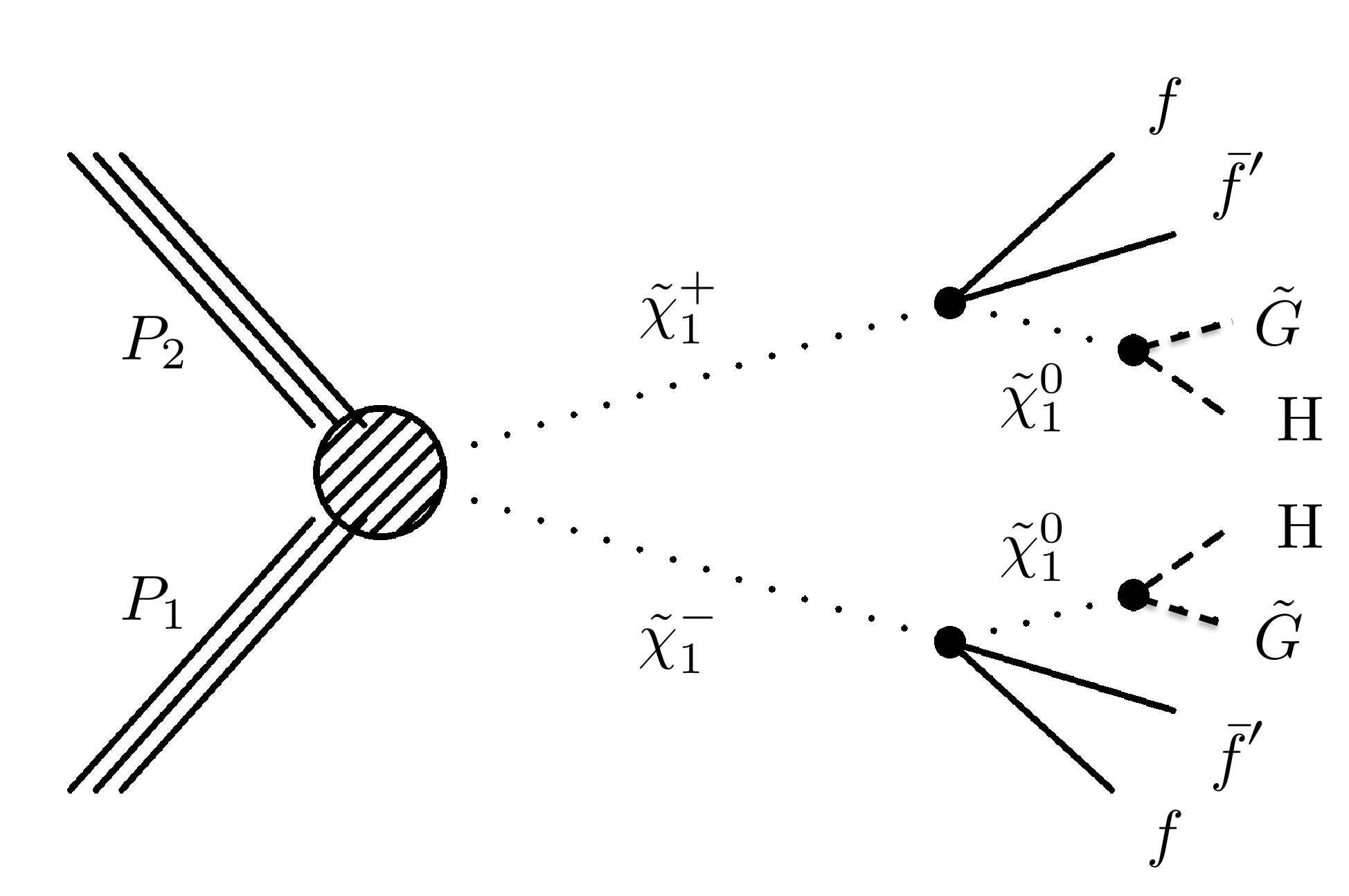}
    \caption{
Example Feynman diagram of strong (\cmsLeft) and electroweak (\cmsRight) higgsino production.
The jets or leptons resulting from transitions between higgsinos are extremely soft in
both diagrams due to the near mass degeneracy among the higgsinos.
    \label{fig:feyn}}
  \end{center}
\end{figure}

The final state of interest, therefore, has two Higgs bosons, some missing energy from the
gravitinos escaping detection, and, in the case of strong production, two bottom (or top, if kinematically allowed) quarks
from top squark decays, as shown in Fig.~\ref{fig:feyn}.
SM backgrounds, \eg from top quark pair production, are suppressed by taking advantage of the known Higgs boson mass and by considering the case when at least one of the Higgs bosons decays into two photons and at least two $\cPqb$-quarks are present from either Higgs or top squark decays.
This approach also allows us to use diphoton mass sidebands for an
estimate of the background  from data, without
sensitivity to the exact composition of the background,
which is dominated by QCD production of $\Pgg\Pgg \bbbar$ events and
$\Pgg \bbbar + j$ events with the jet being misidentified as a photon.

We use a data sample of 19.7\fbinv collected during the LHC 8\TeV running with the CMS detector, which is described in detail elsewhere~\cite{cms-detector}.
The detector comprises a silicon pixel and strip tracker covering the pseudorapidity range $\abs{\eta}<2.4$
(where $\eta = -\ln[\tan(\theta / 2)]$ and $\theta$ is the polar angle of the particle with
respect to the counterclockwise proton beam direction) and electromagnetic and hadronic
calorimeters, which cover $\abs{\eta}<3.0$ and $\abs{\eta}<5.0$, respectively. The tracking and calorimeter
systems are housed inside a 3.8 T solenoidal magnet. Muon chambers are imbedded in the iron return yoke
of the solenoid up to pseudorapidity of $\abs{\eta}< 2.4$.
Events are required to pass a suite of diphoton triggers that impose
mild shower shape and isolation requirements and demand transverse
energies ($\ET = E \cdot \sin\theta$)
above 36 and 22\GeV for the leading and the sub-leading photon, respectively.

\begin{figure}[hbtp]
  \begin{center}
    \includegraphics[width=1.1\cmsFigWidth]{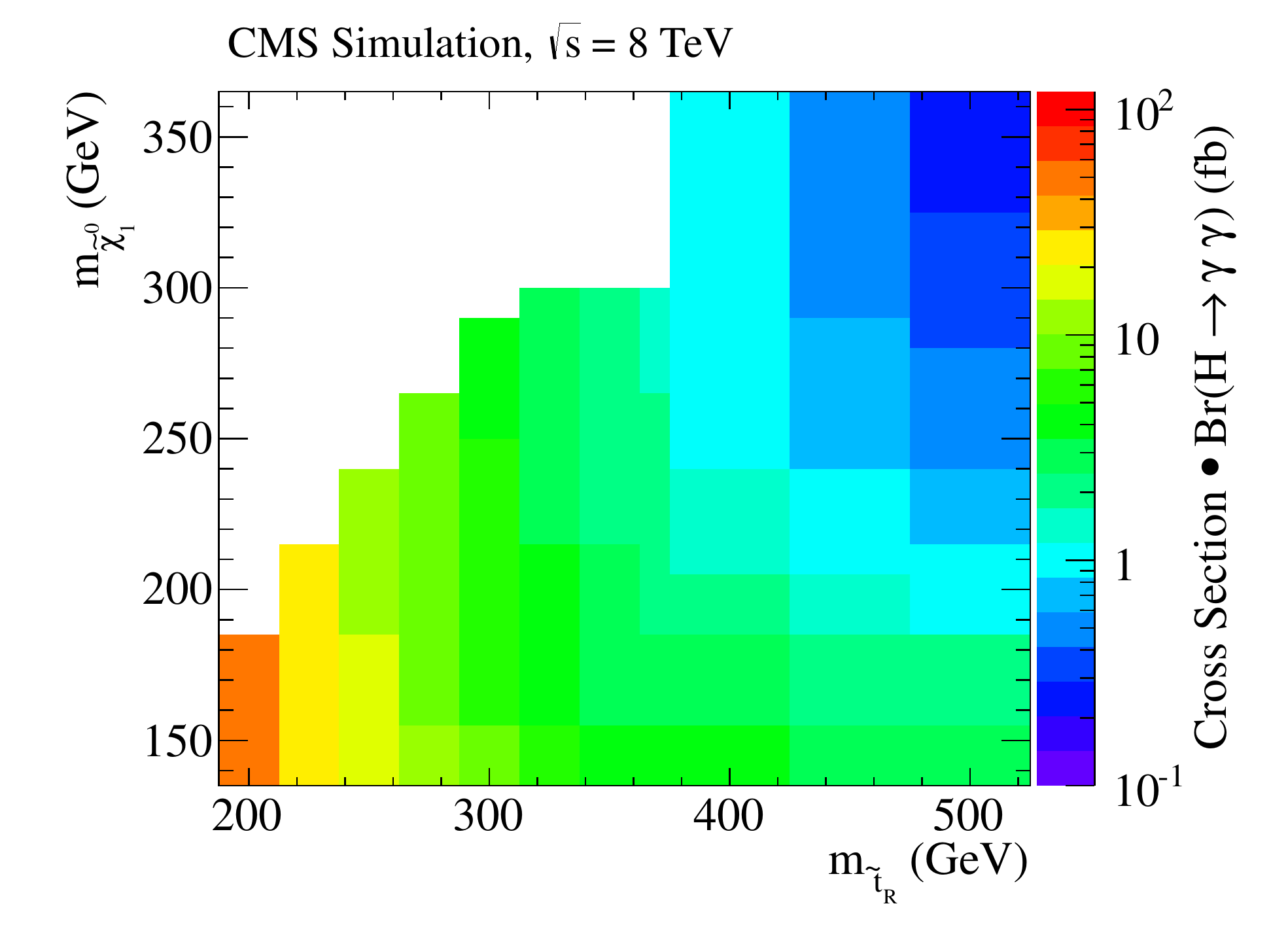}
    \caption{Cross section times branching ratio $\text{Br}(\PH\to\Pgg\Pgg)$ as a function of top squark and higgsino masses.}
    \label{fig:xsec}
  \end{center}
\end{figure}

The signal is simulated for a grid of higgsino and top squark masses
using the \MADGRAPH5 v.1.5.4~\cite{Alwall:2011uj} and \PYTHIA 6.4.22~\cite{pythia3}
generators  and fast simulation of the CMS detector~\cite{Abdullin:1328345}.
The next-to-leading order (NLO) cross sections (shown in Fig.~\ref{fig:xsec})
are calculated using \PROSPINO 2.1~\cite{ref:xsec,NLO1,NLO2,NLO3,NLO4,Beenakker:1996ed}.
The higgsino mass splittings $m_{\PSGczDt}-m_{\conep}$ and $m_{\conep}-m_{\PSGczDo}$ are set to 5\GeV, and a 100\% branching fraction is assumed for the $ \PSGczDo \to \PH \PXXSG $ decay. All other superpartners have masses too heavy to influence SUSY production at 8\TeV.

Photon candidates are reconstructed from the energy deposits in the electromagnetic calorimeter
by grouping the calorimeter's data channels into superclusters~\cite{CMS-PAS-EGM-10-005, cms-egm}.
Photons are required to register in the barrel portion of the calorimeter  ($\abs{\eta} < 1.4442$) and to pass shower shape and isolation requirements.
To reject electrons misidentified as photons, photons are also required not to have associated hit patterns in the pixel detector consistent with a track.
Photon energies are calculated using multivariate regression~\cite{cms-h, cms-egm}.

The particle-flow (PF) algorithm~\cite{cite_MET} is used to reconstruct individual particles
(PF candidates) in the events,
combining all available sub-detector information in a coherent and optimal manner.
Jets are reconstructed from the PF particles using the anti-\kt~\cite{cite_kt} algorithm
with a distance parameter of 0.5
and are required not to overlap with the two leading photons by imposing a requirement
 $\DR > 0.6$, where
 $\DR = \sqrt{(\Delta \phi)^2 + (\Delta \eta)^2}$ and $\phi$ is the azimuthal angle.
Transverse momentum and $\eta$-dependent corrections to the jet energies are applied to account for residual effects of non-uniform
detector response. The pileup contribution is estimated and subtracted from the jet energy using
the jet area method~\cite{Cacciari2008119} on an event-by-event basis.
The combined secondary vertex (CSV) algorithm~\cite{cite_csv} is employed to identify jets that originate from a bottom quark.

The missing transverse energy ($\MET$) of the event is computed as the magnitude of the vectorial sum of
the transverse momenta of all PF candidates~\cite{cite_MET}.

Events are selected if they have at least two identified photons with transverse energies above 40 and 25\GeV for
the leading and sub-leading photons, respectively.
Events are also required to have at least two jets with transverse energy above 30\GeV and $\abs{\eta} < 2.4$
and satisfying the loose (CSV-loose) $\cPqb$-tag requirements, at least one of which also must
 pass the medium (CSV-medium) requirement.
The CSV-loose and CSV-medium working points have signal efficiencies of 80-85\% and 50-75\%,
 respectively, and mistag rates from light-flavor jets of 8-13\% and 1-2\%, respectively \cite{cite_csv}. 

The events with diphoton mass between 120 and 131\GeV constitute the \textit{signal sample},
while the events with mass between 103 and 118\GeV and between 133 and 163\GeV compose the
\textit{lower sideband} and \textit{upper sideband} samples, respectively.

In the investigated SUSY signal, $\cPqb$-quark jets can originate from
decays of the other Higgs boson and from decays of top squark.
This is exploited by separating events into three categories with different expected background as follows:
(i)~events with at least one additional {CSV-loose} $\cPqb$-quark jet in addition to the two (\ie events with three or more $\cPqb$-quark jets),
(ii)~events for which the invariant mass of the two $\cPqb$-quark jets is within a Higgs mass window from 95 to 155\GeV,
(iii)~all other events.

The distribution of signal events among the three categories depends on the
top squark and higgsino masses. For small top squark-higgsino mass differences,
most of the signal populates category (ii), while for large mass differences
categories (i) and (iii) dominate.

The search is performed independently in the three categories and the results are
combined, leading to as much as 35\% improvement in expected SUSY cross section limits
compared to the analysis without categorization.

The peaking background from the SM Higgs boson production with subsequent diphoton decay
is found negligible using MC simulation. The diphoton mass
sidebands are used to derive an estimate of all other SM background processes from data.

The diphoton distribution is fitted using a power law from 103 to 163\GeV \cite{ref:supp}. The region from 118 to 133\GeV,
corresponding to the signal region plus the 2\GeV ``buffers'' on each side of the signal region,
is excluded from the fitter's consideration to prevent potential signal from affecting the background estimate.
The result of the fit is used to calculate the normalization of the background.
The effects of various other fit functions have been explored, and the resulting changes in
the estimated background yields are well within the uncertainty in the fit function
integral from the power law fit.
The fit and background estimates are done independently in all three event categories.

Binned background distributions of relevant kinematic variables are obtained
for each event category separately
from the lower and upper
sidebands and are weighted according to the fit function integrals. This provides two independent estimates
of the distribution of the SM background in the corresponding variable.
No significant correlations are observed between the diphoton mass and other kinematic variables.
However, to account for possible correlations, the background yield is conservatively assumed to be bounded by the independent background estimates from the upper and lower sidebands.
The background prediction using this bound as a systematic uncertainty is calculated as follows.
The main background estimate in each bin of a kinematic variable is formed by taking an average of the two background estimates from the sidebands.
Half the difference between the two estimates is taken as a systematic uncertainty for that bin, so that the central value and its uncertainty spans the range between the two estimates coming from the lower and upper sidebands.
For the main variable used in this analysis, \MET, the background
uncertainty is dominated by the statistical uncertainty in the sideband samples.
The small correlation between the bin errors, originating from the
common normalization from the fit, is taken into account during the statistical analysis described below.

\begin{figure}[h!tb]
  \begin{center}
    \includegraphics[width=0.91\cmsFigWidth]{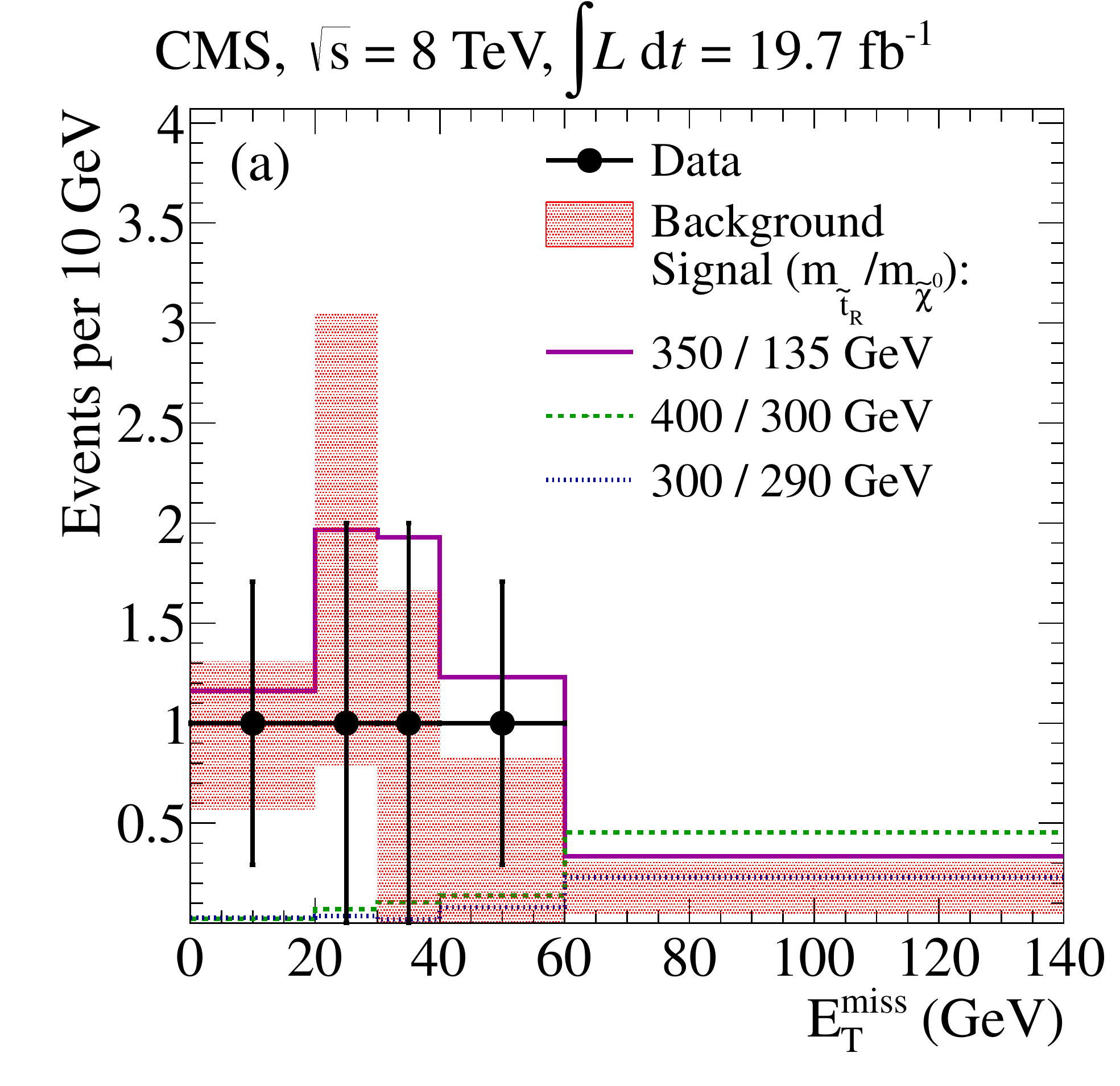}
    \includegraphics[width=0.91\cmsFigWidth]{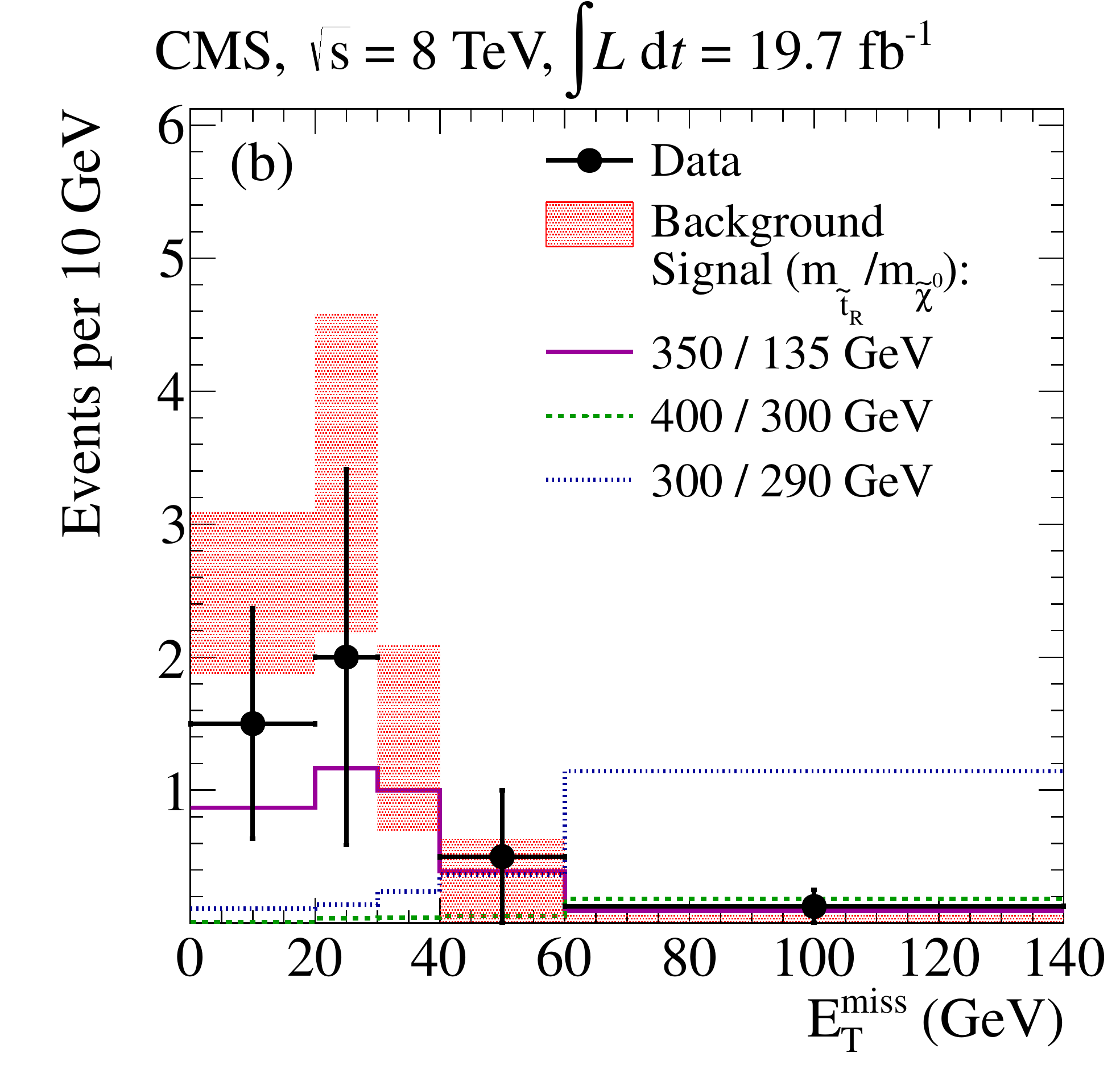}
    \includegraphics[width=0.91\cmsFigWidth]{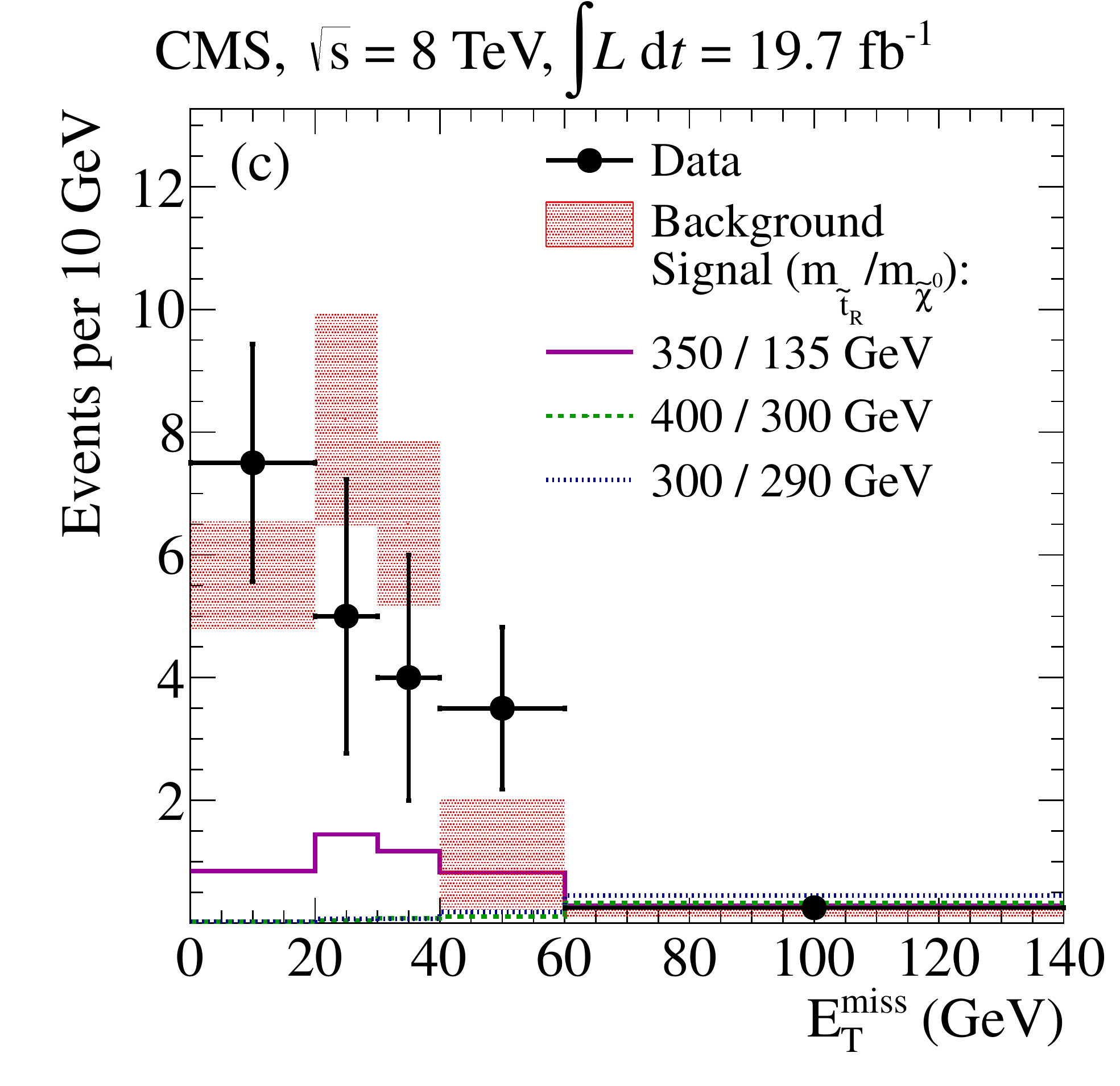}
\caption{\MET distributions for the data, background predictions, and representative signals for
the three event categories (i), (ii), and (iii) in frames (a), (b) and (c), respectively.
For each histogram, the last bin includes the overflow. }
\label{metbbin}
  \end{center}
\end{figure}

The observed distributions of $\MET$
are shown in Fig.~\ref{metbbin} separately for the three event categories,
together with the background estimates from the above procedure and the expected
distributions from three characteristic signal points.
Table~\ref{tab_observed} shows the
expected signal yields,
background predictions, and
observed event counts
for the three event categories.
The observations are in agreement with the background predictions.

\begin{table}[htbH]
\centering
\topcaption{Expected and observed event counts. The yields for three signal points are shown
for comparison indicating $m_{\stR}$ and $m_{\PSGczDo}$ in \GeV, respectively. \label{tab_observed}}
\begin{scotch}{cccc}
 Category   & (i) & (ii) & (iii) \\
\hline
signal 350 / 135  & 10.7 &  2.0 & 6.8 \\
signal 300 / 290  & 2.1  & 10.1 & 3.9  \\
signal 400 / 300  & 4.0  & 1.4 & 2.8   \\
\hline
expected background & 6.7 $\pm$ 1.4 & 10.5 $\pm$ 1.8 & 29.7 $\pm$ 2.8 \\
\hline
observed  & 6 & 7 & 33 \\
\end{scotch}
\end{table}

\begin{figure}[htp]
  \begin{center}
    \includegraphics[width=1.1\cmsFigWidth]{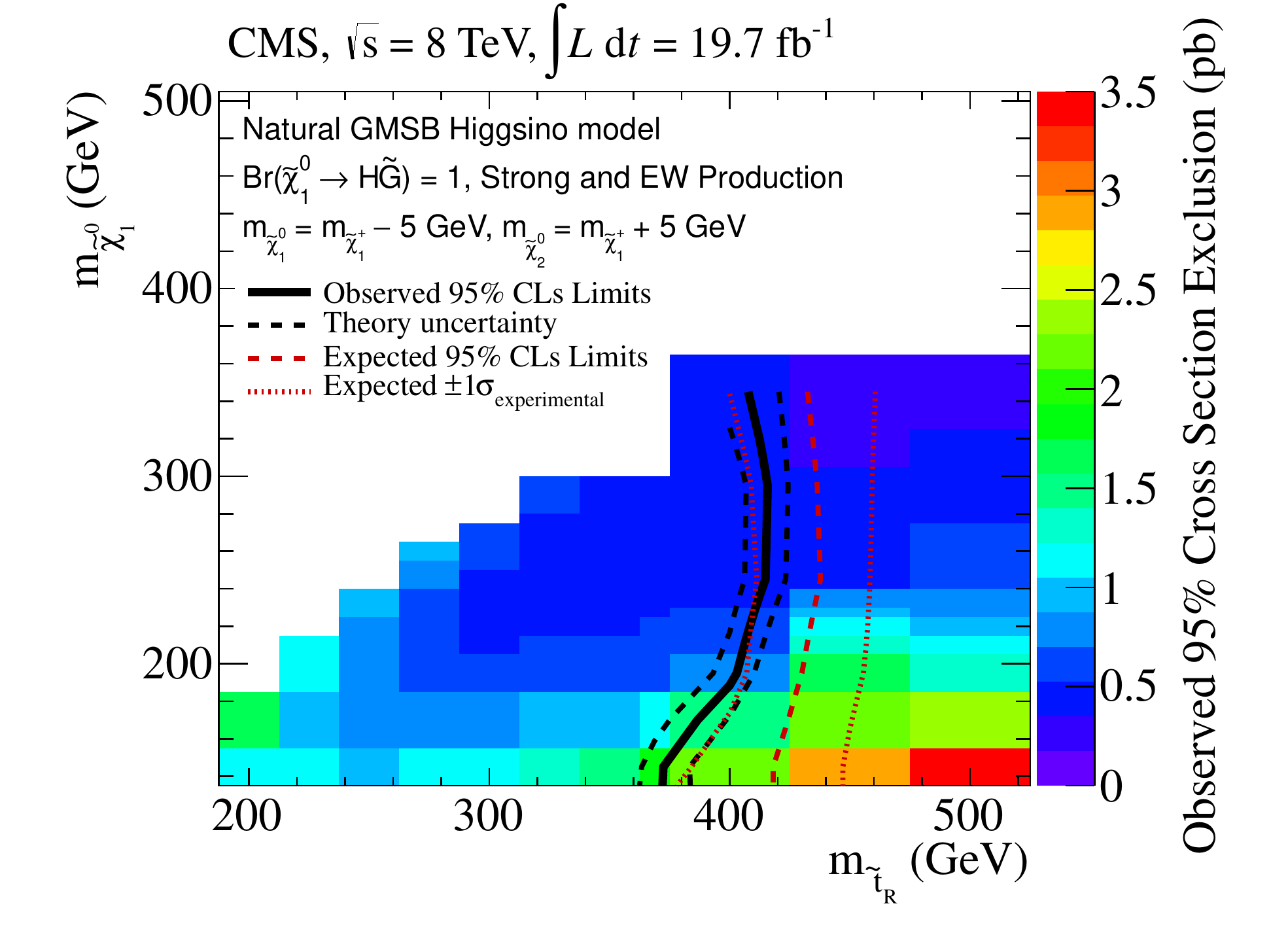}
    \caption{Limits on SUSY production cross sections for different top squark and higgsino masses.
The regions to the left of the contours are expected (thick dash/red) and observed (solid/black) to be excluded at the 95\% confidence level.
Black dashed curves reflect the theoretical uncertainty on the cross section. Red dotted curves correspond to
one standard deviation in expected sensitivity.
}
    \label{fig:limit1}
  \end{center}
\end{figure}

Since the data agree with the expected background we proceed to set limits on the SUSY model described above.
The expected limits were calculated using a variety of kinematic variables and it was determined that \MET is the single most sensitive variable.

The observed \MET distributions for the three event categories,
shown in Fig.~\ref{metbbin}, together with the signal and background expectations
listed in Table~\ref{tab_observed}, are used as input to the limit setting procedure.

A frequentist LHC-style profiled likelihood test statistic is used~\cite{LHlimit,cls1,cls2}. For each model mass point,
the modified frequentist CL$_\mathrm{S}$ method is used to calculate upper
limits on the cross section for the model.

The dominant uncertainty in the analysis is the statistical uncertainty
of the background prediction. The leading sources of systematic uncertainties are
the $\cPqb$-quark jet identification efficiency
and the jet energy scale. The magnitude of the uncertainties depends on the event category and \MET.
For $\cPqb$-quark jet identification it varies from 1 to 5\% for categories with two $\cPqb$-quark jets and from 6 to 17\% for the
three or more $\cPqb$-quark jets.
For the jet energy scale, the uncertainty is important only for \MET values below 40\GeV and
varies from 7 to 43\%.

Other sources of systematic uncertainty in the expected signal yield
 include the uncertainties in integrated luminosity (2.6\%)~\cite{lumiref},
the diphoton trigger efficiency (0.1\%), the photon reconstruction and identification efficiency (1\%),
and the photon resolution uncertainty (1\%)~\cite{cms-h}.
All systematic uncertainties are correlated among the event categories and are
treated as nuisance parameters in the likelihood, profiled according to their
estimated value.

Fig.~\ref{fig:limit1} shows the limits on the GMSB model in the top squark-higgsino mass plane.
Depending on the higgsino mass,
and conservatively using the minus-one-standard-deviation value for the theoretical cross section \cite{ref:xsec},
top squark masses below 360 to 410\GeV are excluded at a 95\% confidence level, corresponding 
to the region to the left of the leftmost thin dashed black line in Fig.~\ref{fig:limit1}. 

To summarize, a search for a ``natural" SUSY scenario with light higgsinos and a top squark was performed, using Higgs tagging in the diphoton decay mode,
in a final state containing at least two photons and two or more $\cPqb$-quark jets.
No evidence for a signal is observed, and 95\% confidence level limits are set in the top squark-higgsino mass plane, excluding a significant
portion of the ``natural" parameter space up to top squark masses of 410\GeV.

We congratulate our colleagues in the CERN accelerator departments for the excellent performance of the LHC
and thank the technical and administrative staffs at CERN and at other CMS institutes for their contributions
to the success of the CMS effort. In addition, we gratefully acknowledge the computing centers and personnel
of the Worldwide LHC Computing Grid for delivering so effectively the computing infrastructure essential to
our analyses. Finally, we acknowledge the enduring support for the construction and operation of the LHC and
the CMS detector provided by the following funding agencies: BMWF and FWF (Austria); FNRS and FWO (Belgium);
 CNPq, CAPES, FAPERJ, and FAPESP (Brazil); MES (Bulgaria); CERN; CAS, MoST, and NSFC (China); COLCIENCIAS (Colombia);
MSES (Croatia); RPF (Cyprus); MoER, SF0690030s09 and ERDF (Estonia); Academy of Finland, MEC, and HIP (Finland);
 CEA and CNRS/IN2P3 (France); BMBF, DFG, and HGF (Germany); GSRT (Greece); OTKA and NIH (Hungary); DAE and DST (India);
 IPM (Iran); SFI (Ireland); INFN (Italy); NRF and WCU (Republic of Korea); LAS (Lithuania);
CINVESTAV, CONACYT, SEP, and UASLP-FAI (Mexico); MBIE (New Zealand); PAEC (Pakistan); MSHE and NSC (Poland);
FCT (Portugal); JINR (Dubna); MON, RosAtom, RAS and RFBR (Russia); MESTD (Serbia); SEIDI and CPAN (Spain);
Swiss Funding Agencies (Switzerland); NSC (Taipei); ThEPCenter, IPST, STAR and NSTDA (Thailand);
TUBITAK and TAEK (Turkey); NASU (Ukraine); STFC (United Kingdom); DOE and NSF (USA).

\bibliography{auto_generated}   

\providecommand{\href}[2]{#2}\begingroup\raggedright\begin{thebibliography}{10}%
\makeatletter
\providecommand{\hrefCMSnoop }[0]{\@secondoftwo}%
\makeatother
\providecommand{\doi}{\texttt{doi:}\begingroup \urlstyle{tt}\Url}

\bibitem{atlas-hcoup}
\hrefCMSnoop {} {{ ATLAS} Collaboration, ``Measurements of Higgs boson
  production and couplings in diboson final states with the {ATLAS} detector at
  the {LHC}'',} \textit{ Phys. Lett. B} \textbf{ 726} (2013) 88,
  \href{http://dx.doi.org/10.1016/j.physletb.2013.08.010}{\doi{10.1016/j.physletb.2013.08.010}},
  \href{http://www.arXiv.org/abs/1307.1427}{\texttt{ arXiv:1307.1427}}.

\bibitem{atlas-hspin}
\hrefCMSnoop {} {{ ATLAS} Collaboration, ``Evidence for the spin-0 nature of
  the Higgs boson using {ATLAS} data'',} \textit{ Phys. Lett. B} \textbf{ 726}
  (2013) 120,
  \href{http://dx.doi.org/10.1016/j.physletb.2013.08.026}{\doi{10.1016/j.physletb.2013.08.026}},
  \href{http://www.arXiv.org/abs/1307.1432}{\texttt{ arXiv:1307.1432}}.

\bibitem{PhysRevLett.110.081803}
\hrefCMSnoop {} {{ CMS} Collaboration, ``{Study of the Mass and Spin-Parity of
  the Higgs Boson Candidate via its Decays to Z Boson Pairs}'',} \textit{ Phys.
  Rev. Lett.} \textbf{ 110} (2013) 081803,
  \href{http://dx.doi.org/10.1103/PhysRevLett.110.081803}{\doi{10.1103/PhysRevLett.110.081803}},
\href{http://www.arXiv.org/abs/1212.6639}{\texttt{ arXiv:1212.6639}}.

\bibitem{cms-h}
\hrefCMSnoop {} {{ CMS} Collaboration, ``{Observation of a new boson with mass
  near 125 GeV in pp collisions at $\sqrt{s}$ = 7 and 8 TeV}'',} \textit{ JHEP}
  \textbf{ 06} (2013) 081,
  \href{http://dx.doi.org/10.1007/JHEP06(2013)081}{\doi{10.1007/JHEP06(2013)081}},
\href{http://www.arXiv.org/abs/1303.4571}{\texttt{ arXiv:1303.4571}}.

\bibitem{susy-natural-1}
\hrefCMSnoop {} {R.~Barbieri and G.~F. Giudice, ``{Upper Bounds on
  Supersymmetric Particle Masses}'',} \textit{ Nucl. Phys. B} \textbf{ 306}
  (1988) 63,
\href{http://dx.doi.org/10.1016/0550-3213(88)90171-X}{\doi{10.1016/0550-3213(88)90171-X}}.

\bibitem{susy-natural-2}
\hrefCMSnoop {} {M.~Papucci, J.~T. Ruderman, and A.~Weiler, ``{Natural SUSY
  endures}'',} \textit{ JHEP} \textbf{ 09} (2012) 035,
  \href{http://dx.doi.org/10.1007/JHEP09(2012)035}{\doi{10.1007/JHEP09(2012)035}},
\href{http://www.arXiv.org/abs/1110.6926}{\texttt{ arXiv:1110.6926}}.

\bibitem{gmsb-1}
\hrefCMSnoop {} {S.~Dimopoulos, M.~Dine, S.~Raby, and S.~D. Thomas,
  ``{Experimental signatures of low-energy gauge mediated supersymmetry
  breaking}'',} \textit{ Phys. Rev. Lett.} \textbf{ 76} (1996) 3494,
  \href{http://dx.doi.org/10.1103/PhysRevLett.76.3494}{\doi{10.1103/PhysRevLett.76.3494}},
\href{http://www.arXiv.org/abs/hep-ph/9601367}{\texttt{ arXiv:hep-ph/9601367}}.

\bibitem{gmsb-kribs}
S.~Ambrosanio\hrefCMSnoop {} { {et~al.}, ``{Search for supersymmetry with a
  light gravitino at the Fermilab Tevatron and CERN LEP colliders}'',} \textit{
  Phys. Rev. D} \textbf{ 54} (1996) 5395,
  \href{http://dx.doi.org/10.1103/PhysRevD.54.5395}{\doi{10.1103/PhysRevD.54.5395}},
\href{http://www.arXiv.org/abs/hep-ph/9605398}{\texttt{ arXiv:hep-ph/9605398}}.

\bibitem{gmsb-2}
\hrefCMSnoop {} {K.~T. Matchev and S.~D. Thomas, ``{Higgs and $Z$ boson
  signatures of supersymmetry}'',} \textit{ Phys. Rev. D} \textbf{ 62} (2000)
  077702,
  \href{http://dx.doi.org/10.1103/PhysRevD.62.077702}{\doi{10.1103/PhysRevD.62.077702}},
\href{http://www.arXiv.org/abs/hep-ph/9908482}{\texttt{ arXiv:hep-ph/9908482}}.

\bibitem{gmsb-3}
\hrefCMSnoop {} {K.~Howe and P.~Saraswat, ``{Excess Higgs production in
  neutralino decays}'',} \textit{ JHEP} \textbf{ 10} (2012) 065,
  \href{http://dx.doi.org/10.1007/JHEP10(2012)065}{\doi{10.1007/JHEP10(2012)065}},
\href{http://www.arXiv.org/abs/1208.1542}{\texttt{ arXiv:1208.1542}}.

\bibitem{gmsb-shih}
\hrefCMSnoop {} {P.~Meade, M.~Reece, and D.~Shih, ``{Prompt decays of general
  neutralino NLSPs at the Tevatron}'',} \textit{ JHEP} \textbf{ 05} (2010) 105,
  \href{http://dx.doi.org/10.1007/JHEP05(2010)105}{\doi{10.1007/JHEP05(2010)105}},
\href{http://www.arXiv.org/abs/0911.4130}{\texttt{ arXiv:0911.4130}}.

\bibitem{cms-detector}
\hrefCMSnoop {} {{ CMS} Collaboration, ``The {CMS} experiment at the {CERN}
  {LHC}'',} \textit{ JINST} \textbf{ 3} (2008) S08004,
\href{http://dx.doi.org/10.1088/1748-0221/3/08/S08004}{\doi{10.1088/1748-0221/3/08/S08004}}.

\bibitem{Alwall:2011uj}
J.~Alwall\hrefCMSnoop {} { {et~al.}, ``{MadGraph 5: going beyond}'',} \textit{
  JHEP} \textbf{ 06} (2011) 128,
  \href{http://dx.doi.org/10.1007/JHEP06(2011)128}{\doi{10.1007/JHEP06(2011)128}},
\href{http://www.arXiv.org/abs/1106.0522}{\texttt{ arXiv:1106.0522}}.

\bibitem{pythia3}
\hrefCMSnoop {} {T.~Sj{\"o}strand, S.~Mrenna, and P.~Z. Skands, ``{PYTHIA 6.4
  physics and manual}'',} \textit{ JHEP} \textbf{ 05} (2006) 026,
  \href{http://dx.doi.org/10.1088/1126-6708/2006/05/026}{\doi{10.1088/1126-6708/2006/05/026}},
\href{http://www.arXiv.org/abs/hep-ph/0603175}{\texttt{ arXiv:hep-ph/0603175}}.

\bibitem{Abdullin:1328345}
\hrefCMSnoop {} {{ CMS} Collaboration, ``The fast simulation of the {CMS}
  Detector at the {LHC}'',} in \textit{ International Conference on Computing
  in High Energy and Nuclear Physics (CHEP 2010)}.
\newblock \textit{J. Phys.: Conference Series} 331 (2011) 032049.
  \href{http://dx.doi.org/10.1088/1742-6596/331/3/032049}{\doi{10.1088/1742-6596/331/3/032049}}.

\bibitem{ref:xsec}
M.~Kr{\"a}mer\hrefCMSnoop {} { {et~al.}, ``{Supersymmetry production cross
  sections in pp collisions at $\sqrt{s}$ = 7 TeV}'',} (2012).
\href{http://www.arXiv.org/abs/1206.2892}{\texttt{ arXiv:1206.2892}}.

\bibitem{NLO1}
\hrefCMSnoop {} {W.~Beenakker, R.~Hopker, M.~Spira, and P.~M. Zerwas, ``{Squark
  and gluino production at hadron colliders}'',} \textit{ Nucl. Phys. B}
  \textbf{ 492} (1997) 51,
  \href{http://dx.doi.org/10.1016/S0550-3213(97)80027-2}{\doi{10.1016/S0550-3213(97)80027-2}},
\href{http://www.arXiv.org/abs/hep-ph/9610490}{\texttt{ arXiv:hep-ph/9610490}}.

\bibitem{NLO2}
\hrefCMSnoop {} {A.~Kulesza and L.~Motyka, ``{Threshold resummation for
  squark-antisquark and gluino-pair production at the LHC}'',} \textit{ Phys.
  Rev. Lett.} \textbf{ 102} (2009) 111802,
  \href{http://dx.doi.org/10.1103/PhysRevLett.102.111802}{\doi{10.1103/PhysRevLett.102.111802}},
\href{http://www.arXiv.org/abs/0807.2405}{\texttt{ arXiv:0807.2405}}.

\bibitem{NLO3}
W.~Beenakker\hrefCMSnoop {} { {et~al.}, ``{Soft-gluon resummation for squark
  and gluino hadroproduction}'',} \textit{ JHEP} \textbf{ 12} (2009) 041,
  \href{http://dx.doi.org/10.1088/1126-6708/2009/12/041}{\doi{10.1088/1126-6708/2009/12/041}},
\href{http://www.arXiv.org/abs/0909.4418}{\texttt{ arXiv:0909.4418}}.

\bibitem{NLO4}
W.~Beenakker\hrefCMSnoop {} { {et~al.}, ``{Squark and Gluino
  Hadroproduction}'',} \textit{ Int. J. Mod. Phys. A} \textbf{ 26} (2011) 2637,
  \href{http://dx.doi.org/10.1142/S0217751X11053560}{\doi{10.1142/S0217751X11053560}},
\href{http://www.arXiv.org/abs/1105.1110}{\texttt{ arXiv:1105.1110}}.

\bibitem{Beenakker:1996ed}
\hrefCMSnoop {} {W.~Beenakker, R.~Hopker, and M.~Spira, ``{PROSPINO: A Program
  for the production of supersymmetric particles in next-to-leading order
  QCD}'',} (1996).
\href{http://www.arXiv.org/abs/hep-ph/9611232}{\texttt{ arXiv:hep-ph/9611232}}.

\bibitem{CMS-PAS-EGM-10-005}
\href {http://cdsweb.cern.ch/record/1279143} {{ CMS} Collaboration, ``Photon
  reconstruction and identification at $\sqrt{s}$ = 7 {TeV}'',} CMS Physics
  Analysis Summary CMS-PAS-EGM-10-005, 2010.

\bibitem{cms-egm}
\hrefCMSnoop {} {{ CMS} Collaboration, ``{Energy calibration and resolution of
  the CMS electromagnetic calorimeter in pp collisions at $\sqrt{s} = 7$
  TeV}'',} \textit{ JINST} \textbf{ 8} (2013) 09009,
  \href{http://dx.doi.org/10.1088/1748-0221/8/09/P09009}{\doi{10.1088/1748-0221/8/09/P09009}},
\href{http://www.arXiv.org/abs/1306.2016}{\texttt{ arXiv:1306.2016}}.

\bibitem{cite_MET}
\href {http://cdsweb.cern.ch/record/1279341} {{ CMS} Collaboration,
  ``Commissioning of the Particle-Flow Reconstruction in Minimum-Bias and Jet
  Events from {\Pp\Pp} Collisions at 7 {TeV}'',} CMS Physics Analysis Summary
  CMS-PAS-PFT-10-002, 2010.

\bibitem{cite_kt}
\hrefCMSnoop {} {M.~Cacciari, G.~P. Salam, and G.~Soyez, ``The anti-$k_t$ jet
  clustering algorithm'',} \textit{ JHEP} \textbf{ 04} (2008) 063,
  \href{http://dx.doi.org/10.1088/1126-6708/2008/04/063}{\doi{10.1088/1126-6708/2008/04/063}},
  \href{http://www.arXiv.org/abs/0802.1189}{\texttt{ arXiv:0802.1189}}.

\bibitem{Cacciari2008119}
\hrefCMSnoop {} {M.~Cacciari and G.~P. Salam, ``Pileup subtraction using jet
  areas'',} \textit{ Phys. Lett. B} \textbf{ 659} (2008), no.~1–2, 119,
  \href{http://dx.doi.org/10.1016/j.physletb.2007.09.077}{\doi{10.1016/j.physletb.2007.09.077}},
  \href{http://www.arXiv.org/abs/0707.1378}{\texttt{ arXiv:0707.1378}}.

\bibitem{cite_csv}
\hrefCMSnoop {} {{ CMS} Collaboration, ``{Identification of b-quark jets with
  the CMS experiment}'',} \textit{ JINST} \textbf{ 8} (2013) P04013,
  \href{http://dx.doi.org/10.1088/1748-0221/8/04/P04013}{\doi{10.1088/1748-0221/8/04/P04013}},
\href{http://www.arXiv.org/abs/1211.4462}{\texttt{ arXiv:1211.4462}}.

\bibitem{ref:supp}
\hrefCMSnoop {} {{CMS Collaboration}, ``{Search for top squark and higgsino
  production using diphoton Higgs boson decays}'',} (2013). The fits of
  diphoton mass distributions and distributions of considered kinematic
  variables are shown in \suppMaterial.

\bibitem{LHlimit}
\href {http://cds.cern.ch/record/1379837} {{ATLAS Collaboration, CMS
  Collaboration, and LHC Higgs Combination Group}, ``{Procedure for the LHC
  Higgs boson search combination in Summer 2011}'',} Technical Report
  CMS-NOTE-2011-005. ATL-PHYS-PUB-2011-11, CERN, Geneva, Aug, 2011.

\bibitem{cls1}
\hrefCMSnoop {} {T.~Junk, ``{Confidence level computation for combining
  searches with small statistics}'',} \textit{ Nucl. Instrum. Meth. A} \textbf{
  434} (1999) 435,
  \href{http://dx.doi.org/10.1016/S0168-9002(99)00498-2}{\doi{10.1016/S0168-9002(99)00498-2}},
\href{http://www.arXiv.org/abs/hep-ex/9902006}{\texttt{ arXiv:hep-ex/9902006}}.

\bibitem{cls2}
\hrefCMSnoop {} {A.~L. Read, ``{Presentation of search results: The $CL_s$
  technique}'',} \textit{ J. Phys. G} \textbf{ 28} (2002) 2693,
\href{http://dx.doi.org/10.1088/0954-3899/28/10/313}{\doi{10.1088/0954-3899/28/10/313}}.

\bibitem{lumiref}
\href {http://cms-physics.web.cern.ch/cms-physics/public/LUM-13-001-pas.pdf} {{
  CMS} Collaboration, ``CMS Luminosity Based on Pixel Cluster Counting - Summer
  2013 Update'',} CMS Physics Analysis Summary CMS-PAS-LUM-13-001, 2013.

\end{thebibliography}\endgroup

\ifthenelse{\boolean{cms@external}}{}{
\clearpage
\appendix
\section{Diphoton mass fits}

Figures \ref{mfit1}, \ref{mfit2}, and \ref{mfit3} show the diphoton mass distributions in the 
three event categories and the fits to a power law.

\begin{figure}[hbtp]
  \begin{center}
    \includegraphics[width=0.8\textwidth]{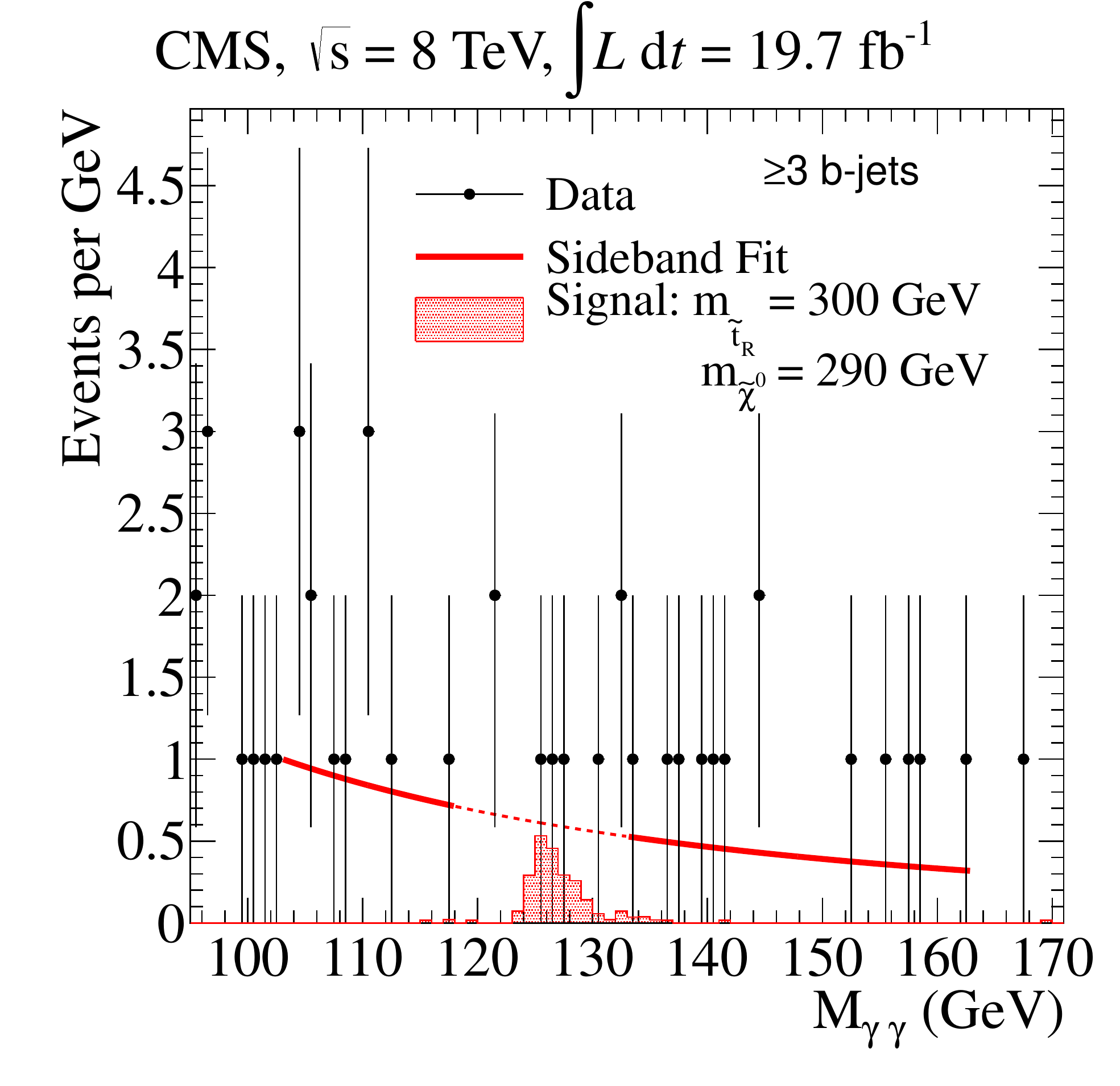} 
    \caption{Diphoton mass distribution (points with errors) in event category (i)
      and the power law fit (thick red line). The interval from 118 to 133 GeV is excluded from the fit.
    \label{mfit1}}
  \end{center}
\end{figure}
\begin{figure}[hbtp]
  \begin{center}
    \includegraphics[width=0.8\textwidth]{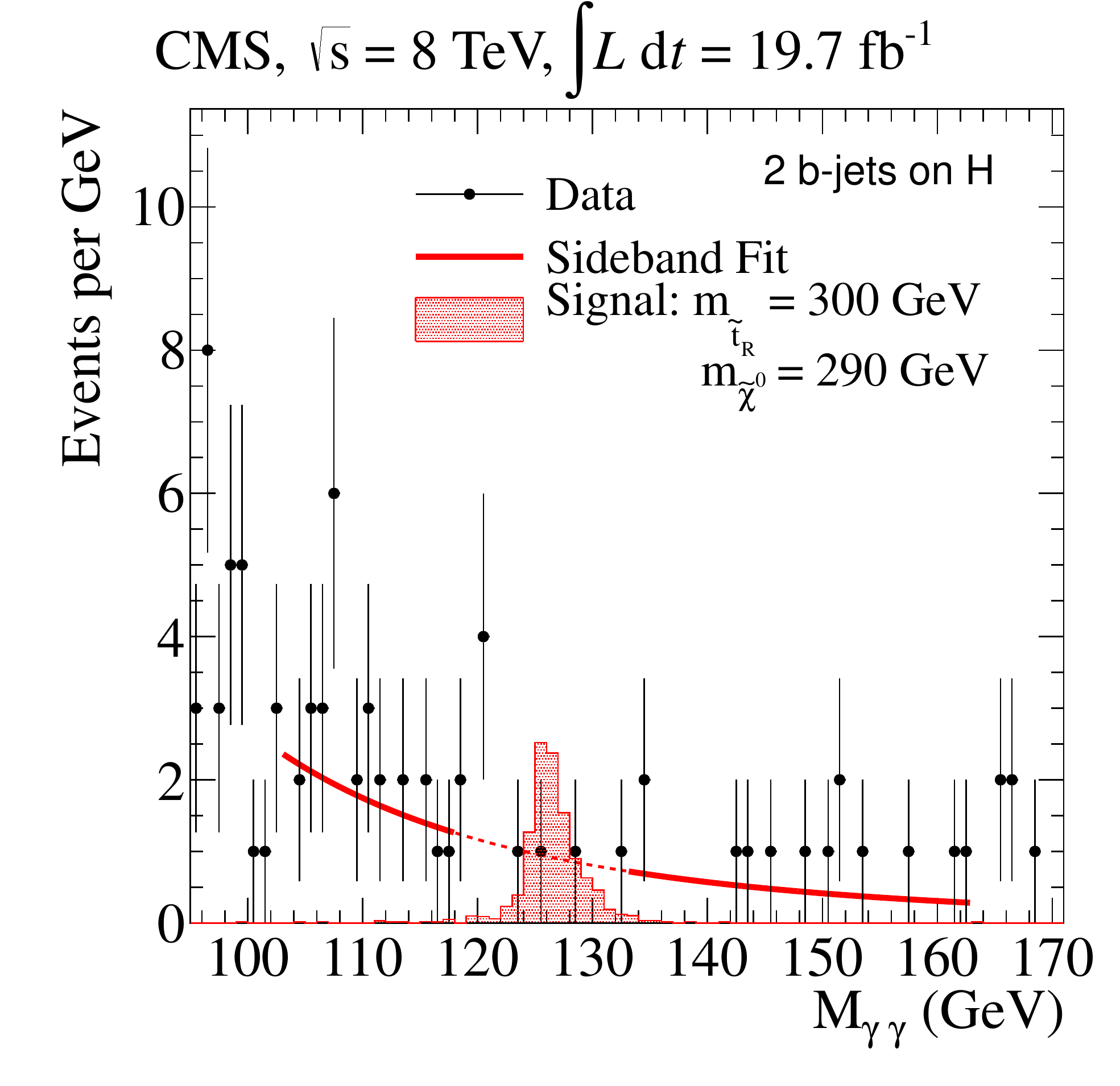} 
    \caption{Diphoton mass distribution (points with errors) in event category (ii)
      and the power law fit (thick red line). The interval from 118 to 133 GeV is excluded from the fit.
    \label{mfit2}}
  \end{center}
\end{figure}
\begin{figure}[hbtp]
  \begin{center}
    \includegraphics[width=0.8\textwidth]{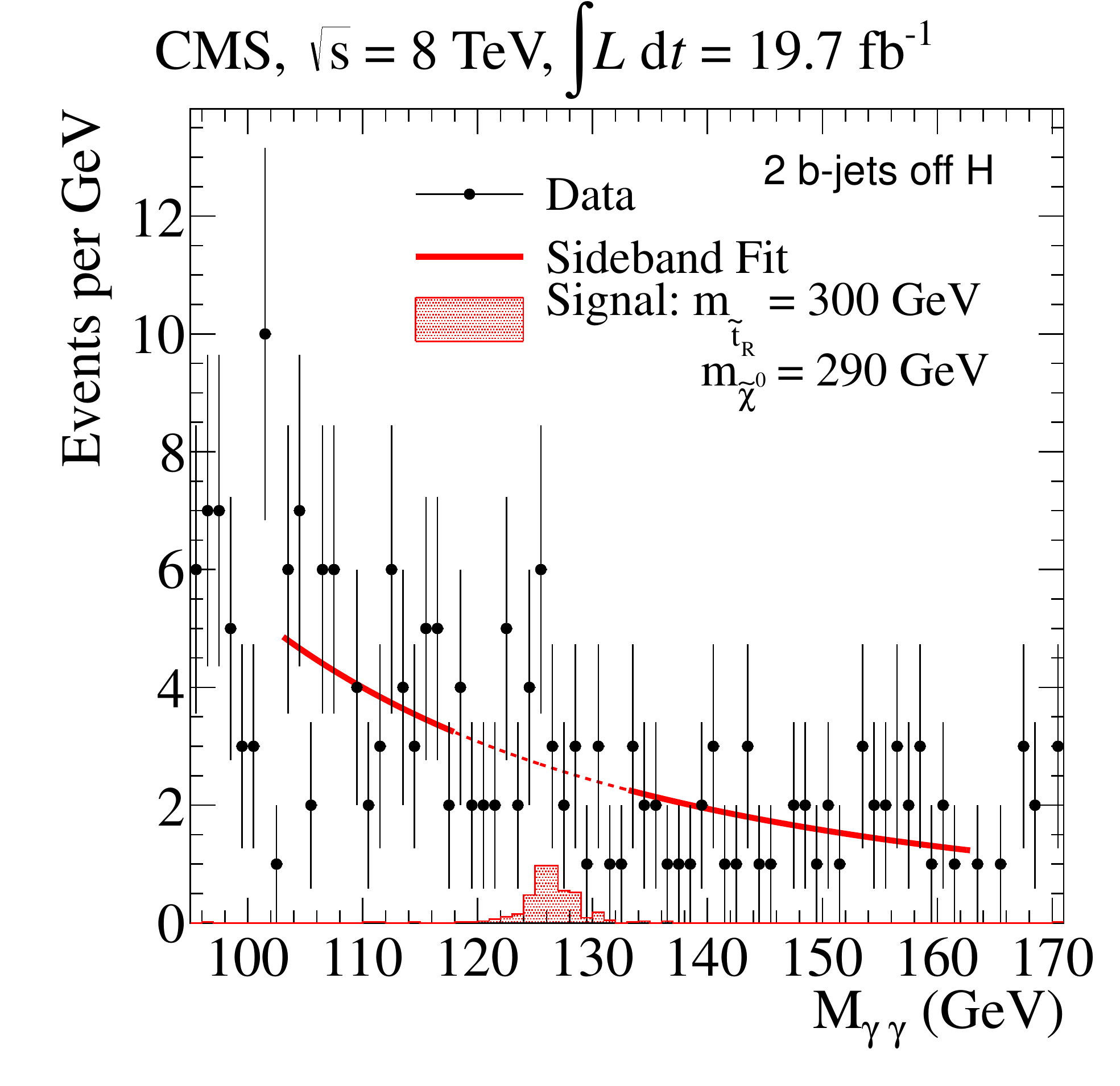} 
    \caption{Diphoton mass distribution (points with errors) in event category (iii)
      and the power law fit (thick red line). The interval from 118 to 133 GeV is excluded from the fit.
    \label{mfit3}}
  \end{center}
\end{figure}

\section{Other kinematic variables}

In addition to \MET, the following kinematic variables were studied:
\begin{itemize}
\item $S_T$, defined as a scalar sum of $p_T$ of all jets and photons in the event;
\item $H_T$, defined as a scalar sum of $p_T$ of all jets in the event;
\item $B_T$, defined as a scalar sum of $p_T$ of all $\cPqb$-quark jets in the event;
\item $H_T^{miss}$, defined as an absolute value of a vector sum of $\vec{p_T}$ of all jets in the event;
\item $p_T^{\gamma\gamma}$, defined as a $p_T$ of the Higgs boson canidate.
\end{itemize}

Figures \ref{extra_st}, \ref{extra_ht}, \ref{extra_mht}, \ref{extra_bt}, and 
\ref{extra_ptgg} show the expected and observed 
distributions of these variables for the three event categories combined, together with expected signal contribution.

\begin{figure}[hbtp]
  \begin{center}
    \includegraphics[width=0.8\textwidth]{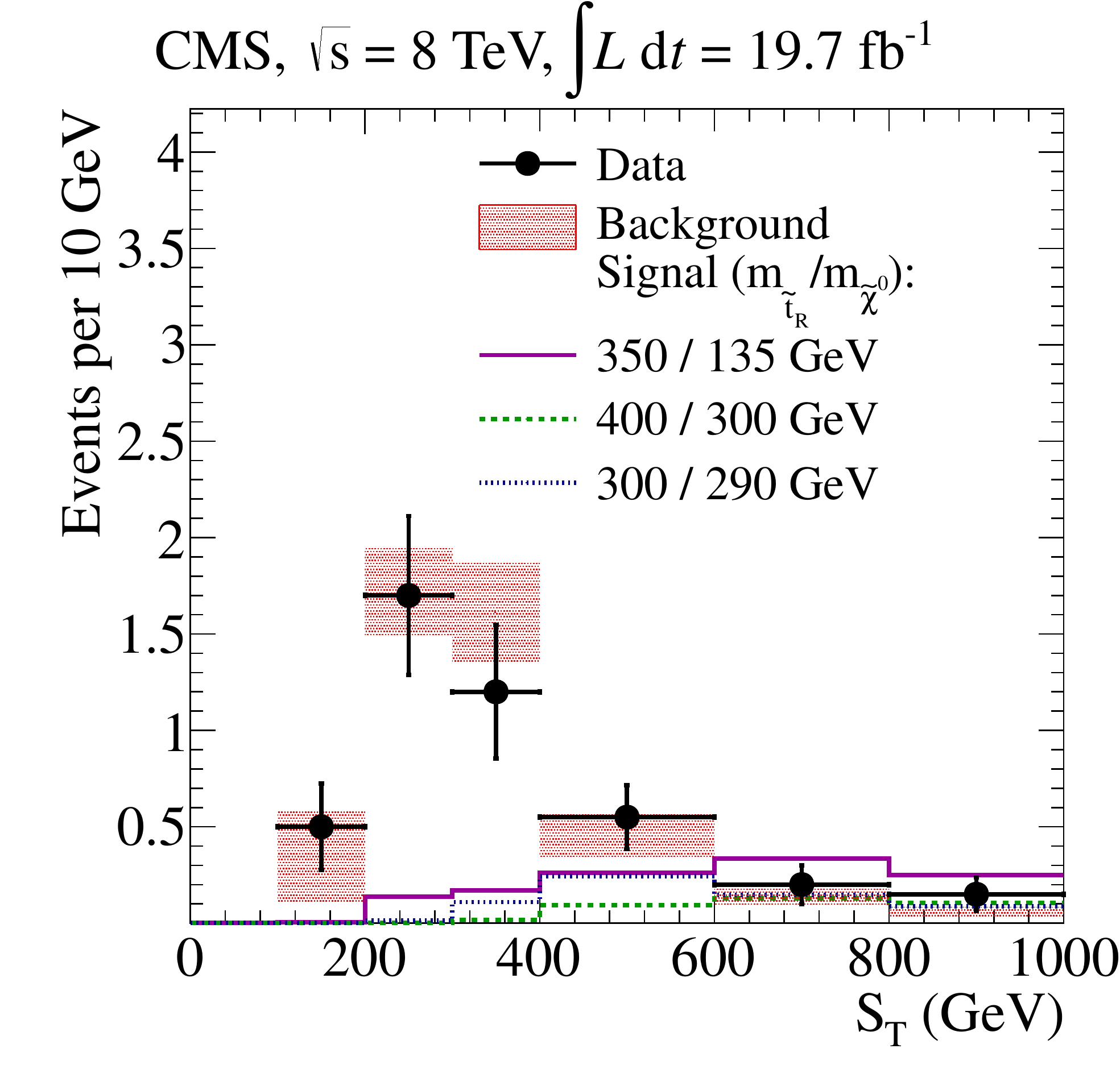} 
    \caption{ $S_T$ distributions for the data, background prediction, and representative signals for
                     the three event categories combined. The last bin includes the overflow.
    \label{extra_st}}
  \end{center}
\end{figure}
\begin{figure}[hbtp]
  \begin{center}
    \includegraphics[width=0.8\textwidth]{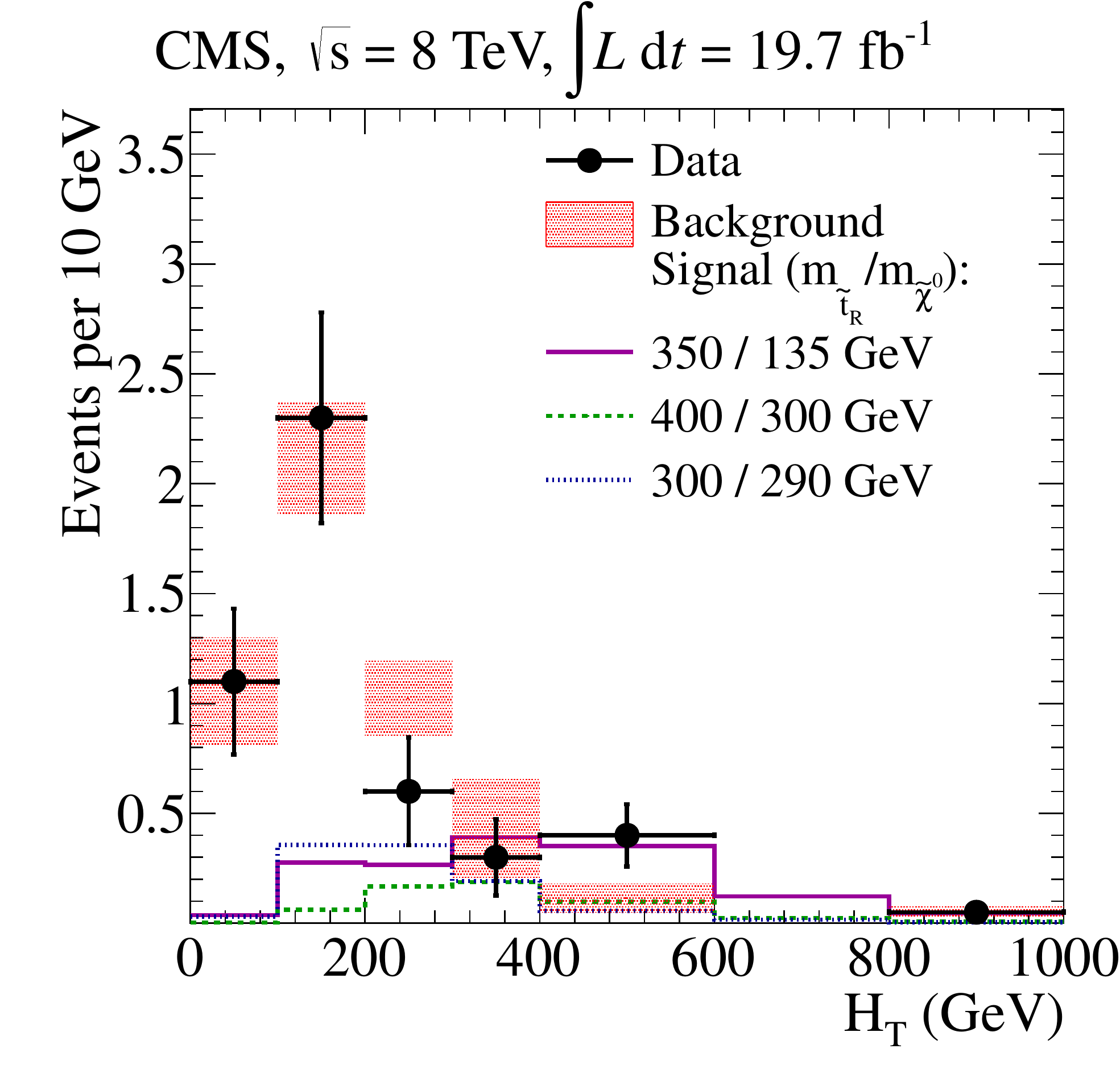} 
    \caption{ $H_T$ distributions for the data, background prediction, and representative signals for
                     the three event categories combined. The last bin includes the overflow.
    \label{extra_ht}}
  \end{center}
\end{figure}
\begin{figure}[hbtp]
  \begin{center}
    \includegraphics[width=0.8\textwidth]{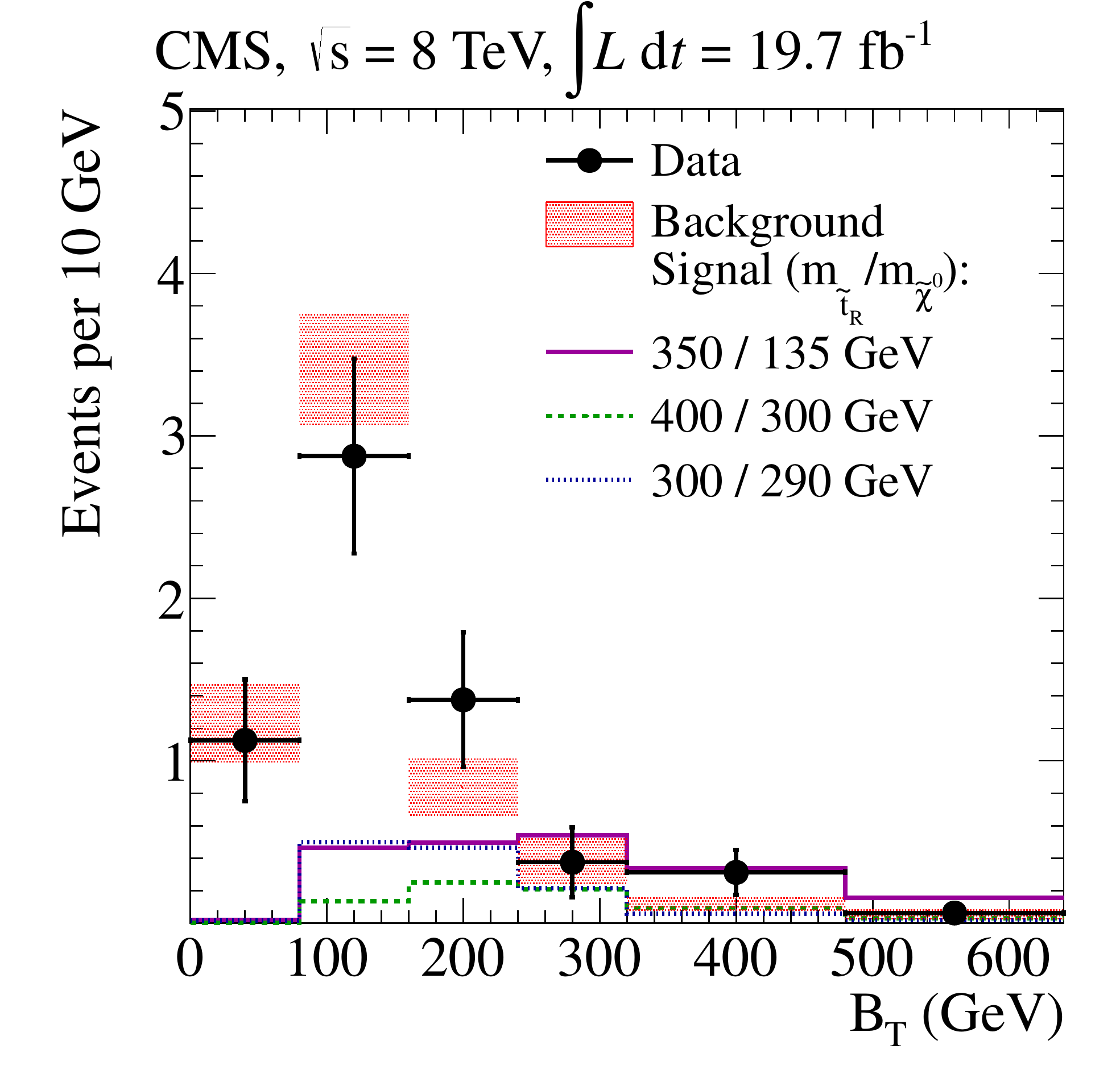} 
    \caption{ $B_T$ distributions for the data, background prediction, and representative signals for
                     the three event categories combined. The last bin includes the overflow.
    \label{extra_bt}}
  \end{center}
\end{figure}
\begin{figure}[hbtp]
  \begin{center}
    \includegraphics[width=0.8\textwidth]{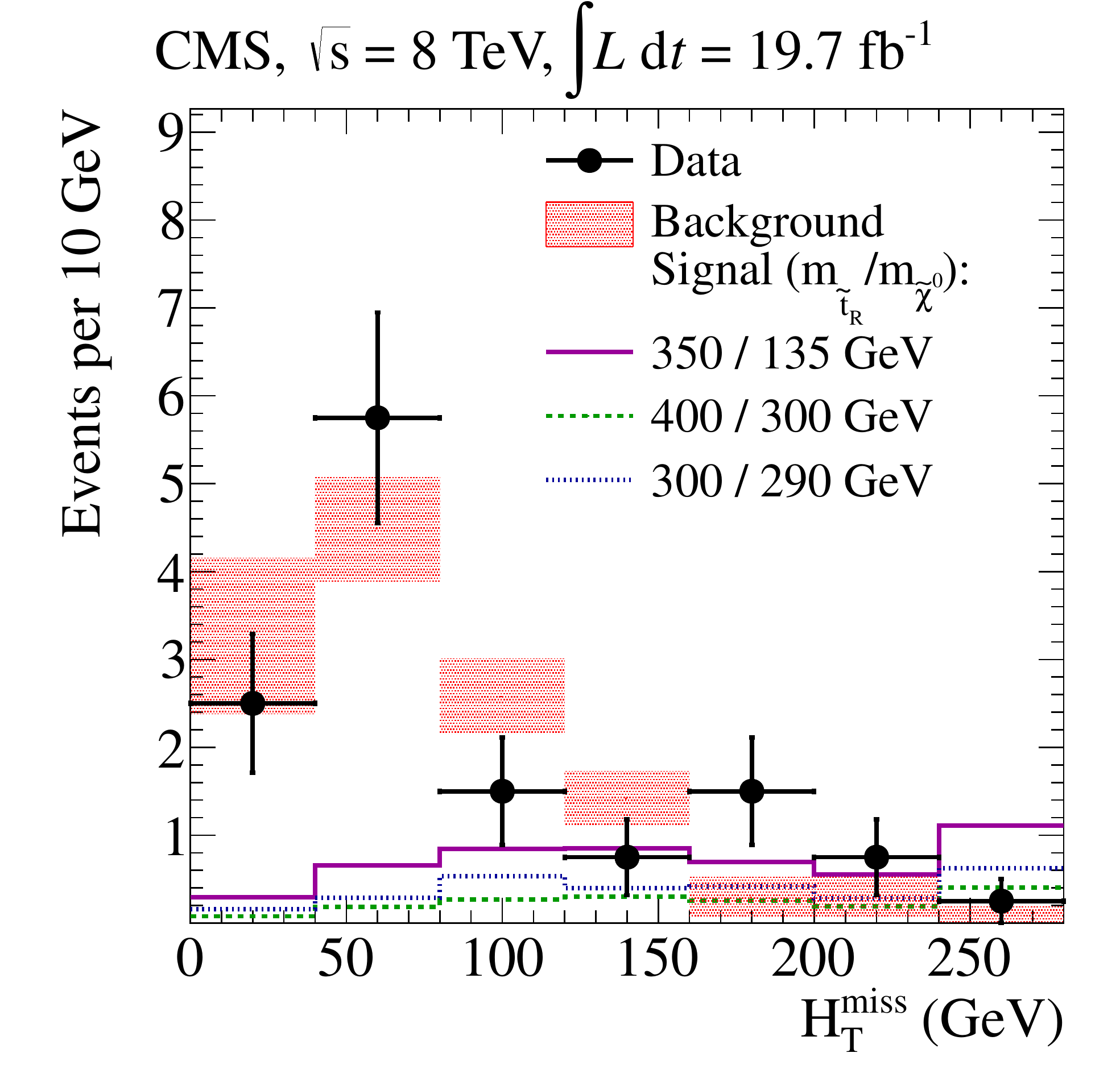} 
    \caption{ $H_T^{miss}$ distributions for the data, background prediction, and representative signals for
                     the three event categories combined. The last bin includes the overflow.
    \label{extra_mht}}
  \end{center}
\end{figure}
\begin{figure}[hbtp]
  \begin{center}
    \includegraphics[width=0.8\textwidth]{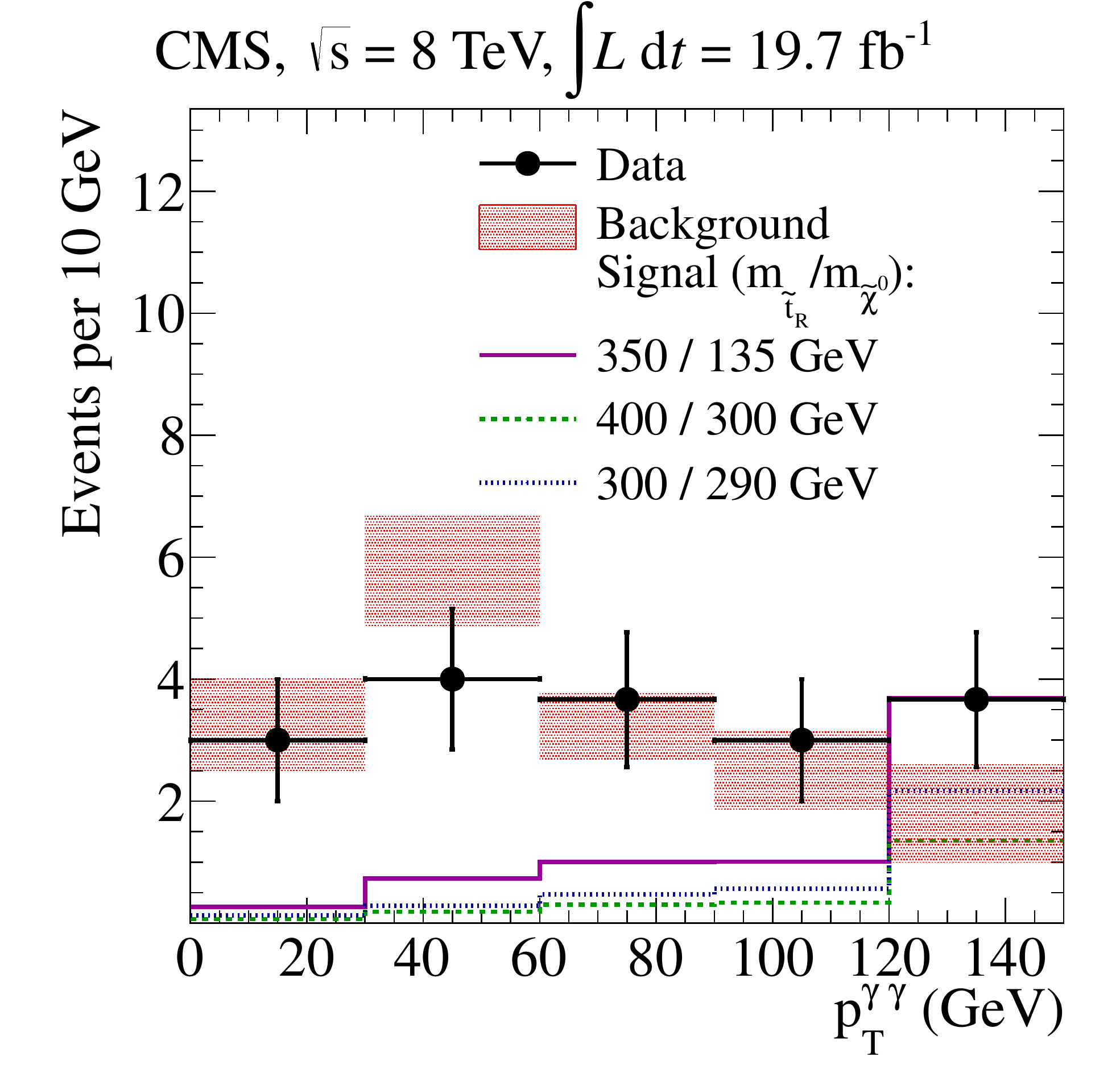} 
    \caption{ $p_T^{\gamma\gamma}$ distributions for the data, background prediction, and representative signals for
                     the three event categories combined. The last bin includes the overflow.
    \label{extra_ptgg}}
  \end{center}
\end{figure}

}

\cleardoublepage \section{The CMS Collaboration \label{app:collab}}\begin{sloppypar}\hyphenpenalty=5000\widowpenalty=500\clubpenalty=5000\textbf{Yerevan Physics Institute,  Yerevan,  Armenia}\\*[0pt]
S.~Chatrchyan, V.~Khachatryan, A.M.~Sirunyan, A.~Tumasyan
\vskip\cmsinstskip
\textbf{Institut f\"{u}r Hochenergiephysik der OeAW,  Wien,  Austria}\\*[0pt]
W.~Adam, T.~Bergauer, M.~Dragicevic, J.~Er\"{o}, C.~Fabjan\cmsAuthorMark{1}, M.~Friedl, R.~Fr\"{u}hwirth\cmsAuthorMark{1}, V.M.~Ghete, C.~Hartl, N.~H\"{o}rmann, J.~Hrubec, M.~Jeitler\cmsAuthorMark{1}, W.~Kiesenhofer, V.~Kn\"{u}nz, M.~Krammer\cmsAuthorMark{1}, I.~Kr\"{a}tschmer, D.~Liko, I.~Mikulec, D.~Rabady\cmsAuthorMark{2}, B.~Rahbaran, H.~Rohringer, R.~Sch\"{o}fbeck, J.~Strauss, A.~Taurok, W.~Treberer-Treberspurg, W.~Waltenberger, C.-E.~Wulz\cmsAuthorMark{1}
% ...
\vskip\cmsinstskip
% ...
1:~~Also at Vienna University of Technology, Vienna, Austria\\
2:~~Also at CERN, European Organization for Nuclear Research, Geneva, Switzerland\\
% ...

\end{sloppypar}
\end{document}